\documentclass{sig-alternate}

\usepackage{clrscode}
\usepackage{listings}

\title{Demaq: A Foundation for \\ Declarative XML Message Processing}

\author{Alexander B\"ohm \hspace*{1cm}
  Carl-Christian Kanne \hspace*{1cm} Guido Moerkotte \vspace{.1cm} \\
Department of Mathematics and Computer Science, University of Mannheim, Germany \vspace{.1cm} \\
{\em alex$|$cc$|$moer}@pi3.informatik.uni-mannheim.de}

\begin{document}
\conferenceinfo{$3^{rd}$ Biennial Conference on Innovative Data Systems Research (CIDR)}{January 7-10, 2007, Asilomar, California, USA.}

\maketitle  

\unitlength1cm

\lstset{basicstyle=\scriptsize, keywordstyle=\color{black}\bfseries\underbar, identifierstyle=, commentstyle=\color{white}, stringstyle=\ttfamily, showstringspaces=true} 
\lstset{emph={fixed, default, value, inherited, in, let, rule, queue, slice, slicing, if, then, return, do, enqueue, create, kind, WSDL, for, where, else, valid, not, define, sliceclass, or, key, and, using, interface, policy, with, qs, message, slicekey, correlated, into, port, on, property, reset, mode, collection, errorqueue, as}, emphstyle=\textbf}
\lstset{fontadjust=yes, basewidth=0.5em}

\begin{abstract}
This paper gives an overview of Demaq, an XML message processing system operating on the foundation of transactional
XML message queues.
We focus on the syntax and semantics of its fully declarative, rule-based application language and demonstrate our message-based
programming para\-digm in the context of a case study. Further, we discuss optimization opportunities for executing Demaq
programs.
\end{abstract}

\section{Introduction} \label{sec:intro}

  The Web is rapidly developing from a one-way medium into an active distributed
  system, where the participating nodes asynchronously communicate via XML
  messages.  Examples for "Active Web" protocols include event notification
  using RSS/Atom feeds \cite{Nottingham:2005:ASF},  business process automation using Web Services \cite{Andrews:2003:BPE, Gudgin:2003:SOA}, and even
  new end-user interface architectures such as AJAX \cite{Garrett:2005:ANA}. Industry sectors as diverse as
  securities trading \cite{FPL:2006:FIX} and multi-media news distribution \cite{IPTC:2006:NAN} have successfully 
  introduced XML messaging as the foundation of their processes. 

Today's systems usually implement these protocols as an
additional tier on top of existing middleware solutions \cite{Alonso:2004:WSC}, further
aggravating the problem of complexity and poor integration that already plagues
these systems~\cite{Stonebraker:2002:TM}: Typically, the actual business
processes are specified using imperative, high-level languages such as Java,
C\# or C++. An incoming message travels through the various
layers: The XML body of the message  is transformed into the
middleware's representation,  again transformed into the programming language's
representation, with further transformations thrown in as other components such
as relational DBMSs are accessed.   Delivering a result requires a reverse
traversal of this "transformation chain".  This not only hurts performance, but
also reduces developer productivity because  each layer requires at least some
separate design and coding that is not related to the actual application
domain. Asynchronous operation and dependability requirements
add more dependencies and complexity. The interaction of various configuration options and code fragments is
difficult to understand, optimize, and maintain.

  The "Demaq" (DEclarative Messaging And Queuing) pro\-ject described here
investigates an alternative approach for the specification and implementation
of such systems.  Our work is based on a simple model: Essentially, the
processes in the whole network as well as the processes on the individual nodes
can be represented as a set of XML message queues and a set of rules for message
exchange between these queues.  The behavior of any node (or group of nodes) can
be completely specified by enumerating its queues and associated rules.

The core idea of Demaq is to use a fully declarative, executable rule language for specification and
implementation of Active Web nodes.  
This allows to move the responsibility for
implementation details from the programmer to the processing system. This
increases productivity if the typical asynchronous "dequeue-process-react"
processing model for message-driven applications becomes part of the language
semantics. Declarativity also facilitates data independence, which in the case
of message processing means that aspects such as message persistence and
recovery, message retention, streamed or materialized representation, or
transport protocols are transparent to the programmer unless their control is
explicitly desired. These degrees of freedom can be used by the processing
system to automatically optimize the execution of the application.  Last but not least, declarativity
simplifies reasoning about the
properties of the system~\cite{Deutsch:2006:VCD}, both on the level of individual nodes and
whole systems.

  We want to answer the question whether such a fully declarative language for XML
message processing is viable from a systems perspective. Our Demaq server 
realizes an Active Web node by executing a declarative program. It 
 leverages database technology for
  message processing by using XML data stores for reliable, transactional XML message queues, and
  declarative XML query processing technology for efficient rule evaluation.
This has never been
done before: Existing approaches in this problem space either (1) work on
network layers below the application~\cite{Loo:2006:DNL}, (2) focus on
relational queue persistence \cite{Doraiswamy:2005:RTI,Gawlick:2003:ISO} and do not provide a control language that natively supports
XML, making the implementation of above-mentioned protocols a pain, (3) are
partly imperative (e.g.
\cite{Abiteboul:1999:AVE,Andrews:2003:BPE,Cooney:2005:APL,Florescu:2003:XLP,Wolter:2005:MSS}),
making an automatic optimization difficult, or (4) only consider isolated
subproblems, omitting a specification of a complete system architecture
\cite{Bailey:2005:FXR, Bonifati:2002:PRS, Onose:04:XYS}. 

  \begin{figure}[t] \centering
  \includegraphics[width=\columnwidth]{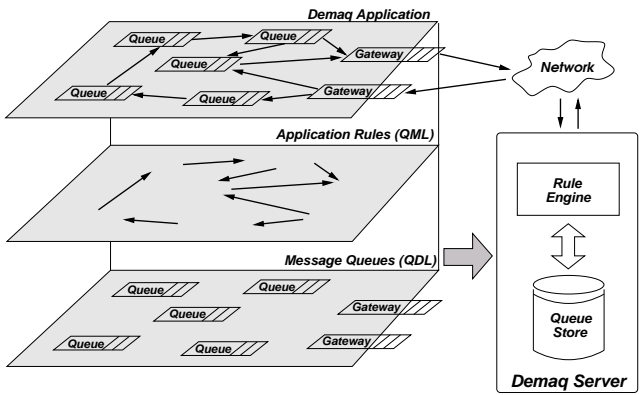}
  \caption{Structure of a Demaq application} 
  \label{arch} \end{figure}

  Our main contribution is
  the fully declarative, executable language Demaq for XML message processing. It is
based on existing and emerging standards such as XML Schema \cite{Thompson:2001:XSS},
XQuery~\cite{Boag:2005:XQL}, 
and the XQuery Update Facility~\cite{Chamberlin:2006:XQU}. The Demaq language has only a few primitives,
  which can be divided into two sublanguages:
  \begin{description}
  \item[A queue definition language] for the specification of
  \begin{itemize}
  \item queues for representing local state,
  \item queues for the communication with remote nodes,
  \item virtual queues, called {\em slices}, which group messages according to user-defined criteria, and
  \item message retention policies. 
  \end{itemize}
  \item[A rule language] for message flow control, which
  \begin{itemize}
  \item is "closed", i.e. describes the system's reaction to messages exclusively in terms of new messages,
  \item extends the XQuery Update language \cite{Chamberlin:2006:XQU} with queuing primitives, and
  \item has simple semantics based on an execution model that suggests a straightforward implementation, but
  leaves room for automatic optimization. 
    \end{itemize}
  \end{description}

  We complement the language description with many examples and a discussion of our ongoing implementation,
  which shows
  \begin{itemize}
  \item why our language can be executed in a reliable, efficient and scalable way,
  \item how existing database technology can be leveraged for message processing (exemplified by our native XML data store Natix \cite{Fiebig:2003:ANX}),
  \item where the opportunities for further research are. 
  \end{itemize}


Fig. \ref{arch} visualizes a Demaq application (top pane). The figure also illustrates
the outline of the paper: An application consists of queue definitions specified using the Queue Definition Language (QDL), including 
communication queues to the outside world (bottom pane and Sec.~\ref{sect:qdl}), and Queue Manipulation Language (QML) application rules for message flow (middle pane and Sec.~\ref{sect:qml}).
Demaq applications are executed by the Demaq server (shown on the right and discussed in Sec.~\ref{sec:impl}). 

\section{Messages and Queues}
\label{sect:qdl}
Demaq applications are based on an infrastructure of physical and logical
structures defined by the Queue Definition Language (QDL). 
Queues provide physical message containers that decouple
message insertion from processing and provide fast message storage and retrieval operations as well as support for reliability and
communication (Sec.~\ref{sec:queues}). We also introduce the notion of {\em
slices}, which are used for creating logical groups of related messages.
We use slices to simplify application development and to
specify how long messages need to be retained physically before they can be deleted (Sec.~\ref{sec:slicing}).

In the QDL definitions in this section, the names of structures are always qualified XML names.
For brevity, we assume the declaration of a default namespace and
omit namespace prefixes.

\subsection{Queues}
\label{sec:queues}
To provide application programs with high performance communication facilities, Demaq incorporates data structures for efficient, asynchronous messaging operations.
For this purpose, message queues have proven their usefulness in a vast number of messaging and integration solutions \cite{Alonso:2004:WSC}. They allow for fast message
handling operations and directly support the asynchronous processing model of Active Web applications. Consequently, queue data structures are used for all
message handling operations in Demaq.

Apart from the simplest solutions, distributed applications have to keep track of their execution state, for example to reflect their current progress with respect to
the business process they implement. Usually, large parts of the state information of an application program is derived from the messages sent to and received from remote communication partners.

One possibility to represent this information is to associate a corresponding runtime context to each application instance.
This approach is used by BPEL \cite{Andrews:2003:BPE} and XL \cite{Florescu:2003:XLP}, where
instance-local variables can be used for storing state information. Contexts that include these
variable bindings have to be kept for each active process instance, which leads to
scalability issues if the number of processes is large. Some execution systems try to overcome this problem by serializing data (\texttt{dehydration}) of "stale" instances.
For example, the Oracle BPEL Process Manager stores application contexts in a relational database system (\texttt{dehydration store}) and reacquires them
when processing continues \cite{Blanvalet:2006:MBP}.

Demaq chooses another approach by requiring all data to be available as XML messages, each of which resides
in exactly one queue. As a result, messages received from remote communication endpoints and internal
state information are handled in a uniform manner, thus simplifying application development. Furthermore, 
by modeling the state of processes as regular data, we can leverage declarative query processing to obtain the relevant data instead of 
constantly loading, manipulating, and saving
 opaque, monolithic runtime contexts.

For reasons of uniformity, system services providing remote communication facilities and timers are also
modeled as message queues. This greatly reduces the number of primitives in the language and makes it easier to understand and use.

\newpage

\subsubsection{Basic Queues}
\textit{Basic queues} provide local message storage facilities. Every queue has a unique name and mode of operation, specifying whether
its content has to be stored persistently or can be transient. The {\tt persistent} queue mode guarantees that in case of a system crash, messages are not lost,
which is important for business processes. \texttt{Transient} queues may be used in those parts of an application that tolerate data loss
or can compensate for it.
The following statement creates a persistent, local queue.

\begin{lstlisting}
create queue finance kind basic mode persistent
\end{lstlisting}

For all queues, there may also be additional, optional parameters, e.g. for specifying a schema all queued messages have to conform to.
Other parameters include a priority level that determines the relative importance of processing
messages from this queue compared to other queues.

\subsubsection{Gateway Queues}
Interaction with remote transport endpoints is a frequent operation for distributed applications such as Web Services.
To provide application programs with a convenient way for remote communication,
\textit{gateway queues} are used to perform messaging operations. They represent local links to
remote endpoints. Messages that are placed into outgoing gateway queues are sent,
while incoming gateway queues contain messages that have been received from other nodes. 
Message properties (see Sec.~\ref{sec:props}) are used to specify recipients and other communication parameters.

The following example creates an outgoing gateway queue to communicate with an external supplier's Web Service. In addition to the queue name and the
kind, we import the supplier's interface definition from a WSDL file and associate some Web Service extensions (WS-ReliableMessaging \cite{Bilorusets:2005:WSR}
and WS-Security \cite{Atkinson:2002:WSS}) with the queue. 

\begin{lstlisting}
create queue supplier kind outgoingGateway mode persistent
  interface supplier.wsdl port CapacityRequestPort
  using WS -ReliableMessaging policy wsrmpol.xml
  using WS -Security policy wssecpol.xml
\end{lstlisting}

Note that in order to use the reliable messaging extensions which 
support reliable sending across system failures, the created queue must be persistent.

From the point of view of the application rules (see below), there is no difference between gateway queues and regular queues.
This also facilitates the distribution of applications over several
nodes by replacing local queues with pairs of gateway queues that connect two sites.

By introducing gateway queues, all network-related operations can be implemented by a communication subsystem providing a queue-based interface.
Sending a message to a remote transport endpoint can be done by inserting it into a corresponding gateway queue.
Unfortunately, distributed architectures such as Web Service applications involve plenty of heterogeneous, independent clients and software layers, each of them being a potential source of
errors \cite{Waldo:1996:NOD}. Thus, communication aspects and network failure notifications cannot be hidden behind a queue-based messaging interface.
Instead, application programs have to become aware of problems encountered by the communication subsystem and implement corresponding error handlers.
We postpone an overview of Demaq's error handling strategy until Sec.~\ref{errorhandling}.

\subsubsection{Time-Based Queues}
\label{sect:timebased}
An important aspect of automated business processes is time. This is true, for example,
in situations where the absence
of action should cause messages to be sent, e.g. for notifications or reminders. 
Similar to network communication, we want to allow timer events without complicating the rule language. Thus, we model
time-based events as message queues. 
An example of such time-based queues are \textit{echo queues}, which
enqueue any message sent to them into some target queue after
a timeout has expired. Both the timeout and target queue are specified as message properties.

The following example creates a persistent echo queue.

\begin{lstlisting}
create queue echoQueue kind echo mode persistent
\end{lstlisting}

\subsection{Message Properties}
\label{sec:props}
Apart from their XML payload, incoming messages are associated with metadata properties such as their size and a message arrival timestamp.
Another kind of metadata is transport protocol information,
such as the initial sender address or connection handles. Connection handles support synchronous communication, where
a response message must be correlated with an existing connection created by an incoming request.
Metadata access has to be possible from application programs, e.g. if an acknowledgment message needs to include the arrival timestamp of the original message. 

A straightforward solution for metadata representation is the semi-structured nature of the XML message format. 
We could encapsulate the application-specific XML payload in a metadata-carrying XML envelope, embedding all metadata information into the message
bodies themselves. 
As a result, both payload and related metadata information were stored uniformly and could be easily accessed by application programs. 

We have chosen a different approach in Demaq because a fully uniform treatment of message body and message metadata  does not take into account
that the access patterns for payload and metadata differ: 
\begin{itemize}
\item Some metadata is maintained by the system and cannot be freely modified. 
\item Things like connection handles should automatically propagate with the messages. 
\item Some metadata may be computable from the message content.
\end{itemize}

Embedding metadata into the message bodies would put the burden of metadata management on the application developer,
who would have to worry about which part of the messages are part of the application specific schema, which parts
are automatically maintained by the system, which parts must be propagated to new messages, etc.
For these reasons, we have included primitives for {\em property} 
management into Demaq.
Properties are key/value pairs, with unique names and a typed, atomic value.  They are determined during message creation and remain fixed
over the message's lifetime. There are several ways to establish property values:

\begin{description} 
\item[Explicit] A property value may be explicitly specified when enqueuing a message (see Sec.~\ref{sec:functions}).
\item[System] Certain properties are set by the system, such as
the name of the rule that created a message, the {time-stamp} at which a message was created,
or the sender of a message for incoming gateway queues.
\item[Inherited] Properties may be {\em inherited}.
A message whose creation was triggered by another message will have the same 
property values as the triggering message for all inherited properties.
\item[Computed] A property value may be computed using an XPath expression
applied to the corresponding message.
\end{description}

As an example, we define a boolean property which
may be used by messages in four different queues. It
is automatically propagated from message to message if not
explicitly set to a different value.

\begin{lstlisting}
  create property isVIPorder as xs:boolean inherited
    queue crm, finance, legal, customer value false
\end{lstlisting}

We may also define properties that have different computed values
based on which queue the message is inserted into, as in the example below.
Here, "orderID" always takes the computed value and may not
be set explicitly (keyword {\bf fixed}). Note that the value
expressions need not be constants, but may be path expressions which
are evaluated against the message body.
This mechanism can be used to give a name to common subexpressions in rules, similar to
views in SQL.

\begin{lstlisting}
  create property orderID as xs:string fixed
    queue order value //orderID 
    queue confirmation value /confirmedOrder/ID 
\end{lstlisting}

\subsection{Slicings}
\label{sec:slicing}
\begin{figure}
\centering
\includegraphics[width=\columnwidth]{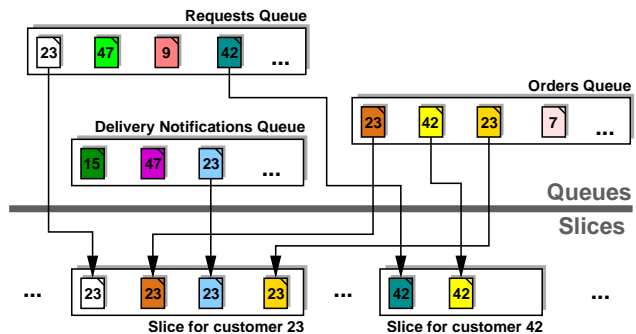}
\label{sstruct}
\caption{Slicing example (customer transactions)}
\end{figure}

Queues represent the most important primitive to physically organize message storage.
However, many applications have multiple, orthogonal criteria for categorizing messages, e.g. all messages belonging to a single business
transaction, all messages received from a particular customer, all priority orders, etc. Typically, messages from several different queues may be part of the same
logical group. This aspect is depicted in Fig. \ref{sstruct}, showing four individual business transactions. Each transaction consists of correlated messages stored in three different,
physical queues (requests, orders, confirmations).

Demaq supports a declarative specification of such logical groups in the form of "virtual" queues,
called {\em slices}. Slices are similar to the concept of parameterized views \cite{Toyama:1986:PVD} for relational databases.
There, a parameterized view defines a family of views, specified by a query expression with a free variable, such that there is one
view relation for each value of the parameter. 
Similarly, a {\em slicing} in Demaq defines a family of virtual queues, where each virtual queue 
consists of all the messages with the same value for a particular property ({\em slice key}).
As rules can be attached to slices, this allows for elegant specification of a variety of typical design
patterns.

\begin{itemize}
\item Slices are a more general form of the "correlation sets" in BPEL \cite{Andrews:2003:BPE} and
"conversations" in XL \cite{Florescu:2003:XLP}.
\item Slices are also useful if the existence of more than one message
is a prerequisite for generating a new message. For example, "joining" parallel control flows
can be implemented by defining a slice, as demonstrated by the example in Sec.~\ref{sliceexample}.
\item Whether or not a message is still required by some process may depend on other messages. Slices allow to group messages to
specify retention policies (see Sec.~\ref{sect:messageretention}).
\end{itemize}

In general, slices are user-defined {\em granularities of data} that are implied by the application domain. 
This not only allows to simplify application specification, but gives rise to many optimization opportunities:
Despite their logical nature, slices can be physically stored 
to speed up message access,  similar to indexes and materialized views. They also
can be used as an additional locking granularity to increase concurrency. 
We will detail
the semantics of slicings in this section and discuss implementation techniques in  Sec.~\ref{sec:impl}. 

\subsubsection{Slicing Definition}
A slicing is created by specifying a unique name and the {\em slicing property},
which may be any property according to Sec.~\ref{sec:props}. 

The property definition lists a number of queues on which the property is defined.
Messages from these queues are partitioned into {\em slices} 
according to their property value. All messages that share the same value of the slicing property are part of
the same slice. 
Each slice represents a "virtual queue" to which rules can be attached (see Sec.~\ref{sliceexample}). 

The property value of a slice is called the {\em slice key}.
The use of property values as slice keys avoids extra language primitives and 
reflects the fact that the ways property values are defined nicely match the 
criteria according to which applications need to group messages: Slice keys sometimes need to be computed 
from the message (using \texttt{computed} property values). In other cases, the rule creating a message might
want to specify the target slice by explicitly setting the slice key. Some messages should
belong to the same slice as the message which caused their creation ({\tt inherited} properties).
The use of system properties such as connection handles enables the application to group all messages 
caused by one particular external message.

In the following example, we group all order and confirmation messages for the
same orderID into a single slice. The slicing "orders" is defined on the property orderID
introduced in Sec.~\ref{sec:props}.

\begin{lstlisting}
  create slicing orders on orderID
\end{lstlisting}

\subsubsection{Slice Resets}
\label{sec:slicereset}

Sometimes, slices have more than one "lifetime". For example, an application
for a domain name registrar may have a slicing based on the domain name to
group all messages related to a particular domain name. If at some point a domain
name changes owners, the application might want to avoid accessing messages 
related to the old owner when using the domain slice.

To indicate that an application is no longer interested in the content of a
particular slice, slices can be "resetted", beginning a new lifetime of the slice.
Only messages of the current lifetime are visible when accessing
the slice. Slightly
more formal, a slice $s$ with a slicekey $k$ contains all those messages that
have a slicing property value of $k$ and have been added after the last reset
operation of $s$.

As slices represent logical groups of messages, resetting a slice does not
necessarily mean the contained messages are physically deleted. We illustrate
how the slicing concept is related to message retention in the following
section.

\subsubsection{Message Retention}
\label{sect:messageretention}

In typical message processing systems, messages are dequeued, i.e. physically removed, from the queue
once they have been processed. However, there are many reasons why messages need to be retained for a
longer period of time. Examples include
legal reasons, auditing, and tracing system behavior. 
In the Demaq model
"everything is a message", and the state of running processes is often encoded in "old" messages scattered throughout the system.
Hence, access to already processed messages
is frequently required. 

This is why Demaq is based on an append-only approach for message queues -- messages are never modified after they
have been created. However, we still need a mechanism for message removal because we cannot assume unlimited
storage capacity.
One straightforward solution is to allow for explicit deletion
by the application program.
This is the equivalent of manual memory management for conventional programming languages, which is a chronic source of errors
and increases
the complexity of development because dependable long-running processes like our Active Web applications need to be free of memory leaks or, in our case,  
"message leaks". Further, explicit deletion leaves no degrees of freedom for the run-time system to optimize execution,
which is inconsistent with our desire for declarativity. In addition, it is often difficult for a single
part of the application code to decide whether a message is not required any more. 
To illustrate this, consider a a procurement application as a simple example for several independent retention criteria: 

\begin{enumerate}
\item From the packaging department's point of view, it is sufficient
to retain an order message until the order has been packed and picked up by a delivery service. Afterwards, the message can be safely deleted. 
\item The corporate finance department requires
the same order message to be retained to make sure that payments received from the customer can be correlated to the corresponding order message.
\item The operations research department requires all order messages to be kept until the end of the month when performing demand planning based on the most
recent customer orders.
\end{enumerate}
When using explicit message deletion primitives in this scenario, the multiple retention requirements cannot be easily combined. In particular,
the order in which the three conditions for safe message deletion become true varies from order to order. Thus, all modules would need to know about the
message retention policy of the other parts of the application, making application maintenance more difficult.

Demaq uses a declarative approach to specify when messages are no longer required.
QML (Sec.~\ref{sect:qml}) does not include an explicit "destroy message" primitive. Instead, the
execution engine only marks whether a message has been processed or not.
This way, physical cleanup is decoupled from message processing and can be done separately, for example in times of low system load or when the remaining storage capacity becomes low.

It turns out that the slicing concept is ideally suited to identify those messages that are still required by the application logic, because
\begin{enumerate}
\item Slices are specified declaratively
\item Slices by design represent contexts in which messages are useful
\item A message may belong to any number of slices
\end{enumerate}   
This is why we chose to couple membership in slices to message retention. The Demaq execution model guarantees that
a message is not physically removed from the message store as long as it is contained in at least one slice. By
resetting slices (Sec.~\ref{sec:slicereset}), applications can indicate that certain "old" messages are no longer interesting
with respect to that slice. Once all slices to which a message belongs have been reset, it is eventually
removed by the system. 
Messages which are not part of any slice may be deleted for the message store as soon as it has been processed.

\begin{figure}
\centering
\includegraphics[width=.8\columnwidth]{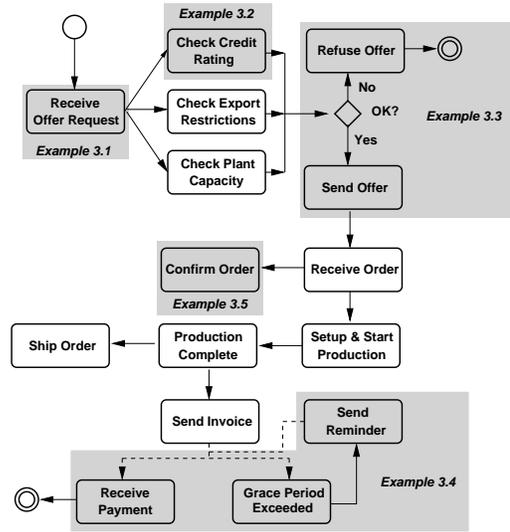}
\caption{Procurement scenario workflow}
\label{exor}
\end{figure}

\section{Queue Manipulation Language}
\label{sect:qml}
The Queue Definition Language introduced in the last section allows the specification of the queue and slicing {infra-structure} used by an Active Web application.
To implement the application logic, it has to be complemented by a convenient programming language. For this purpose, current application servers \cite{Alonso:2004:WSC} successfully
rely on imperative, object-oriented programming languages such as Java, C\#, or Visual Basic.
As these languages do not incorporate XML as a first class data type, auxiliary processing steps are required to convert between the type systems of those languages and the XML format, thus 
harming the overall system performance. Recent efforts, such as Microsoft's LINQ project \cite{Meijer:2006:LIN} or XJ \cite{Harren:2005:XJF}, try to overcome this impedance mismatch.
However, even if XML can be successfully integrated as a first class data type, other disadvantages persist.
For example, the optimization potential of
imperative languages is far lower than that of declarative languages.
Further, it is generally agreed upon that the productivity of programmers is higher
for declarative languages than for imperative ones.

\begin{figure}
\centering
\includegraphics[width=.8\columnwidth]{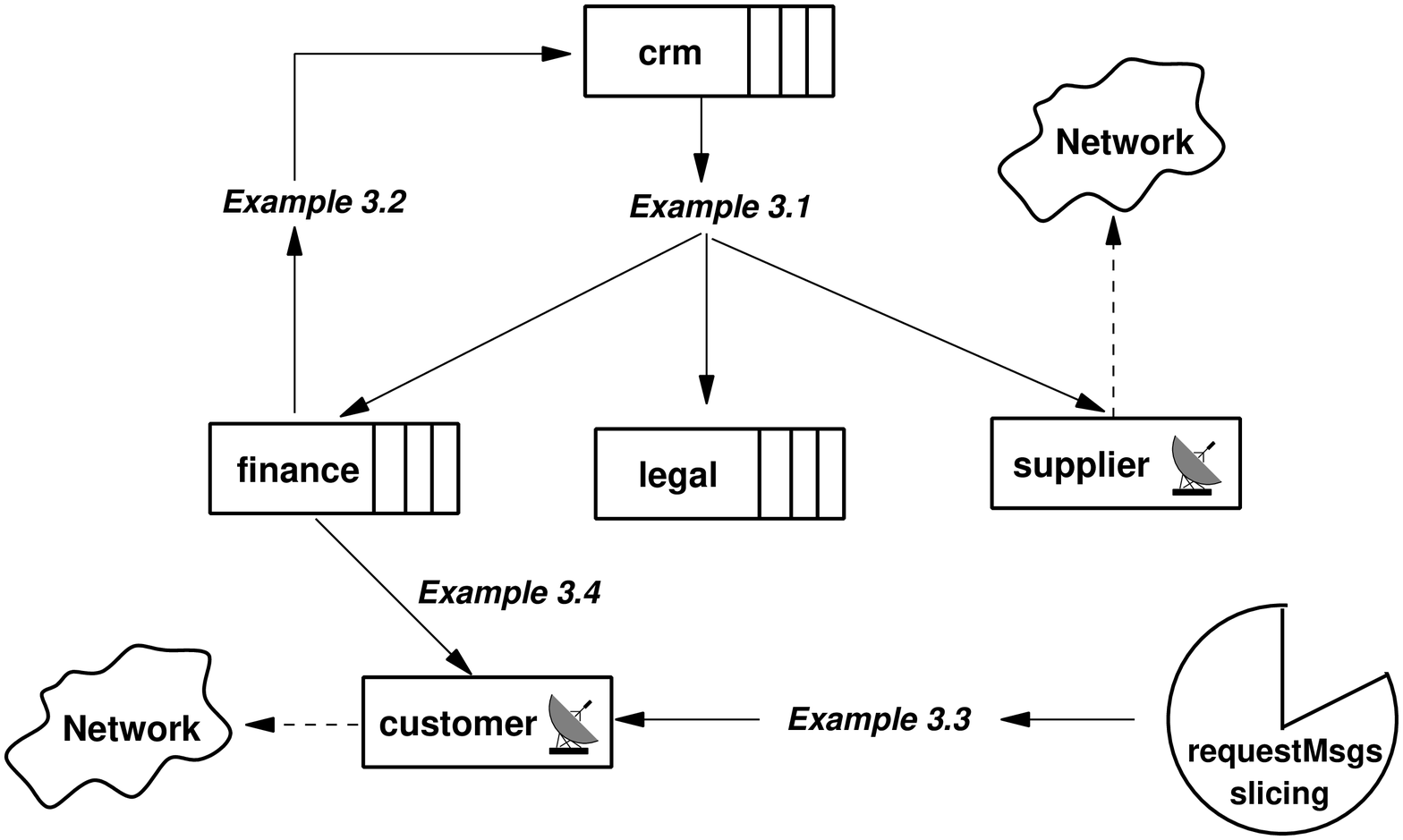}
\caption{Procurement scenario message flow}
\label{mflow}
\end{figure}

To avoid these drawbacks, we choose a declarative language providing native XML support as the basis of our application language.
Our Queue Manipulation Language QML is built on the concept of event-condition-action (ECA) rules that react to a single kind of event,
the arrival of an XML message in a queue.  Each QML rule is essentially an XQuery Update Facility expression \cite{Chamberlin:2006:XQU} attached to a queue. 
Within this section, we first describe the execution model that explains how and when rules are evaluated, before briefly 
recapitulating the features of the XQuery Update Facility. In the remainder of this section, we
develop our QML by implementing parts of an exemplary business process (depicted in Fig. \ref{exor}). This process
is a distributed procurement scenario taken from the chemical industry, where multiple parties, both within the same organizational unit
and on external systems, form a processing network. We provide QML code for some of the steps in the scenario, creating the message flow shown in Fig. \ref{mflow}.

\subsection{Execution Model}
\label{sect:execution}
Following the terminology proposed by Paton and Diaz
\cite{Paton:1999:ADS}, our model applies an iterative cycle policy,
relying on a detached coupling mode that decouples the processing of a message from its
creation. At any given point in time, there may be any number of messages in a node's queues
that have not been processed yet. Each unprocessed message is processed exactly once,
in an order determined by a message scheduler. The scheduler is
influenced by optional priorities declared for the involved queues.

The processing of a message consists of the evaluation of all rules that pertain to
the queue in which the message resides. The evaluation of each rule results in a (possibly empty) list of
pending actions. The resulting actions are state-updating primitives,
with the most frequent action being the creation of a new message and its insertion into another queue.
This list of actions is then executed, and any created messages are made known to the scheduler.
The evaluation of all rules and the subsequent execution of all actions caused by processing a single message
are executed in a single transaction against the message store. Many of such message-processing
transactions may run concurrently to improve performance, as long as they can be executed
in an isolated manner. The separation of rule evaluation from action execution, together with the transaction mapping, ensure a snapshot semantics that facilitates optimization.

\subsection{XQuery Update Facility}
\label{XQUF}
The XQuery Update Facility \cite{Chamberlin:2006:XQU} is intended to allow a declarative specification
of updates in XML data stores.
The bulk of the proposed update language is made up by XQuery 1.0.
In addition, new primitives ({\tt do} statements) are introduced which can be used similar to constructors
to represent pending intra-document updates. Expressions which return such pending updates are called "updating expressions"
and can be combined using existing XQuery constructs such as FLWOR expressions.
Updating expressions produce a pending update list of update primitives that are applied after the entire statement has been evaluated, thus resulting
in a snapshot semantics for expression evaluation.

Using the XQuery Update Facility as the basis for QML has several advantages:
\begin{enumerate}
\item Application developers can benefit from previous programming experience with XQuery.
\item Obviously, the XQuery language is well equipped to process XML data and to construct new XML fragments/messages.
\item The implementation of Demaq becomes easier, as we can reuse existing XQuery implementations.
\end{enumerate}

\subsection{Rule Creation}
\label{sec:ruledef}
The top-level construct in QML is a rule definition.
Each rule has a name (RName) and associates an updating XQuery Update expression with a physical message queue or a slicing (QName). 

\begin{lstlisting}
create rule RName for QName CondExpr
\end{lstlisting}
The body of the rule is always a conditional expression (hence the CondExpr nonterminal) to
facilitate the detection and optimization of conditions by the rule compiler. The CondExpr must
always be an updating expression, making use of the novel queue {\tt do} primitives explained below.
For programming convenience, we allow the "else" part of the CondExpr to be absent and assume that such a rule produces an empty update list in the else case.

\subsection{Message Queue Primitives}
\label{sec:functions}

The result of a rule is a list of pending actions, as explained above.
The most important action resulting from a rule is the creation of a new message and its insertion into a particular queue.
To perform this operation, we extend the list of XQuery update's expressions with a new \texttt{enqueue} update primitive

\begin{lstlisting}
do enqueue ExprSingle into QName
   (with PropName value ExprSingle)*
\end{lstlisting}

which causes a message to be enqueued to the specified queue. The optional {\tt with} clause allows to 
explicitly set properties of the new message (see Sec.~\ref{sec:props}).

\textbf{Example 3.1:}
The QML rule in Fig. \ref{code1} demonstrates how basic message handling and parallelism are performed as a response to the reception of a new customer request
at the customer relationship management (crm) queue of our running example (Fig. \ref{exor}). It initiates three subsequent checks by sending messages to other queues.

The default evaluation context of all XQuery and XPath statements in a QML rule is the document root of the triggering message, thus
''//offerRequest'' matches all offerRequest elements in the incoming message. In this example, we create 
the content of three new XML messages
using \texttt{let}, including the initial request ID and customer ID for message correlation. 
As depicted by Fig. \ref{mflow}, these messages are sent to the "finance", "legal" and "supplier" queues using the \texttt{enqueue} update action, forking the control flow.
To allow message correlation by the supplier's service, 
we add the ''Sender'' property field of the corresponding message using the \texttt{with-value} statement. This metadata information
is automatically interpreted by Demaq's communication subsystem.

\begin{figure}
\begin{lstlisting}
create rule newOfferRequest for crm
if (//offerRequest) then
  let $customerInfo :=
      <requestCustomerInfo>
         {//requestID} {//customerID}
      </requestCustomerInfo>
  let $exportRestrictionInfo := ...
  let $plantCapacityInfo := ...
  return do enqueue $customerInfo into finance,
         do enqueue $exportRestrictionsInfo into legal,
         do enqueue $plantCapacityInfo into supplier
            with Sender value "http://ws.chem.invalid/"
\end{lstlisting}
\caption{Message handling and content access}
\label{code1}
\end{figure}

The QML features a small library of functions (designated by the namespace prefix {\tt qs:}) to access messages and queues from QML rules.
The document node of the currently processed message is returned by \texttt{qs:message()}.
Access to the document nodes of all messages in a particular queue is provided by (\texttt{qs:queue(name)}).
Message properties can be obtained using the \texttt{qs:property(pname)} function, returning the value associated with the "pname" key.

\textbf{Example 3.2}
The rule in Fig. \ref{code2} demonstrates how to formulate predicates which involve both the current message and other messages in a particular queue.
To determine the customers credit rating, the ''invoices'' queue is inspected to find potentially unpaid bills. The messages in the
invoices queue are accessed using the \texttt{qs:queue()} function. To access the content of the triggering message, 
\texttt{qs:message()} can be used (e.g. for correlating the contained customerIDs in the predicate check).

\begin{figure}[htbp]
\begin{lstlisting}
create rule checkCreditRating for finance
if (//requestCustomerInfo)  then
  let $result :=
    <customerInfoResult> {//requestID} {//customerID} 
       {let $invoices := qs:queue("invoices")
        return
          if($invoices[//customerID = qs:message()/customerID])
          then
            <refuse/> (:unpaid bills!:)
          else
            <accept/>}
    </customerInfoResult>
  return do enqueue $result into crm
\end{lstlisting}
\caption{Queue access}
\label{code2}
\end{figure}

\subsection{Slice Primitives}
\subsubsection{Slice Rules}
\label{sliceexample}
Rules can be attached to slicings using the same syntax as for queues. Note
that a slicing specifies many "virtual queues" (slices), and the rule is
attached to every slice of the specified slicing.
\subsubsection{Slice Access}
Like queues, slices are made accessible through additional XQuery functions. The \texttt{qs:slice()} function
returns all messages from the slice of the current message.
The key (property value) of the current slice can be retrieved using the \texttt{qs:slicekey()} function.
Both of these functions are only available to rules defined on slicings, so that they are not ambiguous for messages belonging to more than one slice.

\textbf{Example 3.3}
In our business process (depicted in Fig. \ref{exor}), the decision whether to send or refuse an offer depends on the results of three preceding checks (credit rating, export restrictions and plant capacity), which may run in parallel.
In this example, we use a slicing to join the control flow of the parallel checks before processing continues (Fig. \ref{code3}). It includes
all messages in the crm and customer queues which refer to the given requestID.
\begin{figure}
\begin{lstlisting}
create property requestID as xs:string fixed
  queue crm, customer value //requestID

create slicing requestMsgs on requestID

create rule joinOrder for requestMsgs
if(qs:slice()[/customerInfoResult] and 
   qs:slice()[/restrictionsResult] and
   qs:slice()[/capacityResult]) then 
    if(qs:slice()[/customerInfoResult/accept] and
      not (qs:slice()[/restrictionsResult//restrictedItem])
      and qs:slice()[/capacityResult//accept]) then
      let $request := qs:queue("crm")/offerRequest
      let $items:=$request[//requestID = qs:slicekey()]/items
      let $pricelist := collection("crm")[/pricelist]
      let $offer :=  ...
      return do enqueue $offer into customer 
    else (:problems:)
      do enqueue <refusal>{//requestID}</refusal>
                into customer
\end{lstlisting}
\caption{Control flow synchronization }
\label{code3}
\end{figure}

The rule in Fig. \ref{code3} checks for the arrival of a message in the
slicing and --- if all three preceding checks have been performed --- completes the order request
by sending a reply to the customer.
The \texttt{qs:slice()} function is used to acquire the confirmation messages.
The \texttt{qs:slicekey()} function retrieves the slice key of the current slice for
message correlation with the customer's offer request. Master data (such as
price lists) stored in the non-messaging parts of Demaq is acquired using the
standard XQuery collection() function.

\subsubsection{Slice Resets and Message Retention}
To indicate that the messages in a slice are no longer needed, 
Demaq QML provides a \texttt{reset} update primitive
that resets a slice specified by a slicing name and a slice key. (If no parameters are
given, it resets the slice of the current message with respect to the slicing
to which the current rule is defined.)

\begin{figure}[htbp]
\begin{lstlisting}
create rule cleanupRequest for requestMsgs
  if (qs:slice()/offer or qs:slice()/refusal) then
    do reset
\end{lstlisting}
\caption{Resetting a slice}
\label{code4}
\end{figure}

In the example above, messages of the requestMsgs slice are visible until an offer or a refusal
message with the slice-specific key is finally sent, which completes processing
for this requestID. This causes the "cleanupRequest"
rule to reset the current slice (Fig.~\ref{code4}). Hence, as far as these rules are concerned,
the messages may be deleted after an order request has been processed
completely.  However, there may be other slices which cause the messages to be
retained for longer. Further, note that the slice reset is specified in a separate rule, which causes a reset of
the slice, no matter which rule caused an offer or a refusal to be created. 

When receiving an order, a number of subsequent steps is preformed, including sending order messages to external suppliers, arranging shipment and sending an invoice to the customer until the
business process is successfully terminated by retrieving the payment (Fig. \ref{exor}).
We skip these parts of the process, as they do not provide any additional insight into QML.
However, if the payment has not been received within a particular grace period, a reminder has to be sent to the customer.

\textbf{Example 3.4}
The code snippet in Fig. \ref{code6} shows another example how slicings can be applied to make sure messages are retained.
Here, a QML rule monitors the reception of the timeout notification message previously registered at an echo queue (when sending the invoice to the customer).
If no payment has been received upon expiration of the timeout, it sends a reminder message to the customer. A slicing is used to make sure both invoice and payment confirmation
are retained until the timeout message is received. If the timeout message arrives and the payment has been confirmed, the slice is refreshed by the "resetPayedInvoices" rule.

\begin{figure}
\begin{lstlisting}
create property messageRequestID as xs:string fixed
  queue invoices, finance value //requestID

create slicing invoiceRetention on messageRequestID

create rule resetPayedInvoices for invoiceRetention
if(qs:slice()//timeoutNotification
             and qs:slice()/paymentConfirmation) then
do reset

create rule checkPayment for finance
if (//timeoutNotification) then
  let $mRID := qs:message()//requestID
  let $payments := qs:queue()[/paymentConfirmation]
  return
  if(not($payments[//requestID = mRID])) then
    let $invoice:=qs:queue("invoices")[//requestID = mRID]
    let $reminder := (:access initial invoice:) ...
    return do enqueue $reminder into customer
  else ()
\end{lstlisting}
\vspace{-0.5cm}
\caption{Message retention}
\label{code6}
\end{figure}

\subsection{Error Handling}
\label{errorhandling}

An important aspect of any computer program is its reaction to faults. As the
Demaq engine is supposed to run permanently to guarantee high service
availability, it must be equipped with powerful error handling facilities.
In this section, we review some  sources of errors, and illustrate how they can
be handled in Demaq.

\begin{description}
\item[Application program related]
Some errors are caused by the application program itself. While syntactical
errors or wrong static typing can be detected at compile time and can be handled while developing the
application, there are also many application-related runtime
errors.  For example, as we use XQuery in our rule specifications, processing
might raise one of the various runtime errors defined by the XQuery standard,
which are mainly related to dynamic typing issues.

\item[Message related]
Message related errors may occur when trying to enqueue invalid XML documents received from remote peers into a gateway queue.
For example, input documents may be truncated or not well-formed, thus resulting in parsing errors during processing,
or rules create messages whose schema is incompatible with the target queue's schema.
\item[Network related]
Demaq provides application
developers with a queue-based interface for sending messages over the network by simply enqueuing them into a
corresponding gateway queue. Unfortunately, the existence of the network 
cannot be made completely transparent to application programs. 
Interactions with remote transport endpoints may encounter a variety of network-related issues \cite{Waldo:1996:NOD}.
Among these are low-level network problems such as temporal or permanent unavailability of remote transport endpoints, name resolution failures, timeouts or routing errors.
Further up the protocol stack, there may be other sources of problems. For example, using the WS-Security SOAP enhancements \cite{Atkinson:2002:WSS} may result in
encountering invalid certificates, wrong signatures or decryption failures.

Some of those errors can be resolved by the communication subsystem (message delivery can be automatically retried etc).
However, in many cases, application programs must handle such errors explicitly.

\item[System]
Another source of errors is the Demaq
processing system itself. System failures may occur due to insufficient system resources (such as main memory or secondary storage),
infrastructure problems (e.g. in the underlying database or operating system) or even hardware defects.

\end{description}

To deal with these various kinds of errors, corresponding error handling facilities have to be provided for application developers.
BPEL \cite{Andrews:2003:BPE}, XL \cite{Florescu:2003:XLP} and many other programming languages (including e.g. Java and C++) allow the specification of scoped
exception handlers for those parts of the program that might potentially fail. Extending our expression language XQuery with exception handling would
jeopardize most of the benefits discussed in Sec.~\ref{XQUF}, such as reusing existing XQuery implementations and risk incompatibility with
future versions of XQuery, which might incorporate similar mechanisms.


Instead, like all other events in the Demaq system, errors are represented by {\em XML messages} sent to {\em error queues}. 
If an error is encountered during processing,  a corresponding failure message is inserted
into an error queue. The error message not only contains an error specification according to a predefined schema, but may also contain (a reference to) the data which caused the error,
such as message IDs or corrupt incoming message bodies. Different error queues can be specified at the rule, queue, module and system level.

This way, error handling is integrated seamlessly
with our message and queue-based programming paradigm, reflecting the fact that in many cases, the border between regular application logic 
and error handling is quite fuzzy. Further, we can perform advanced error handling by reusing our regular primitives. For example,
error queues may be gateway queues, notifying remote operators. Error queues may be persistent, guaranteeing eventual reaction to an error 
even in the case of masking higher level (e.g. system) failures.

\textbf{Example 3.5}
The example in Fig. \ref{code7} shows how error handling can be done by a Demaq application. When receiving an customerOrder request, the application rule tries to
send back a confirmation message to the customer. All errors that might be caused by the confirmOrder rule are handled by the crmErrors queue. In this example, an
error handing rule (deadLink) tracks all communication failures caused by disconnected transport endpoints. It compensates for the network failures by sending
a mail confirmation using a corresponding Web Service (connected by the postalService gateway queue).

\begin{figure}[htbp]
\begin{lstlisting}
create queue crmErrors kind basic mode persistent
create property orderID as xs:integer 
  queue crm value //customerOrder/orderID  
create slicing retainOrders on orderID

create rule confirmOrder for crm errorqueue crmErrors
if(//customerOrder) then (:send confirmation:)
let $confirmation := <confirmation>
                      {//orderID} (:additional details:)
                     </confirmation>
return do enqueue $confirmation into customer

create rule deadLink for crmErrors
if (/error/disconnectedTransport) then
  (:send confirmation via snail mail:)
  let $orders := qs:queue("crm")//customerOrders
  let $initialOrderID := /error/initialMessage//orderID
  let $address := $orders[orderID=$initialOrderID]/address
  let $request := <sendMessage>{$address}
     {//initialMessage}</sendMessage>
  return do enqueue $requestMail into postalService

\end{lstlisting}
\caption{Error handling}
\label{code7}
\end{figure}

\section{Implementation Aspects}
\label{sec:impl}

Within this section, we outline some aspects of the ongoing implementation of
Demaq and discuss how certain aspects of the semantics of our declarative language can be used to
optimize execution performance \footnote{Additional information about the Demaq
project is available at http://www.demaq.net/}.

\subsection{Transactional Queues}
\label{sec:taqueues}

Message-oriented middleware solutions successfully rely on transactional queues. Due to their
functional overlap with typical database features such as recovery and security, it has been argued that queues
should be integrated into database systems \cite{Gray:1995:QD}.  Many commercial database vendors have considered
incorporating queues into their products \cite{Doraiswamy:2005:RTI} or already support message queues as native
data structures \cite{Gawlick:2003:ISO, Wolter:2005:MSS}. 

We believe that a queue-enabled XML data store is
the most suitable foundation for the storage subsystem of an active XML message processing system such as Demaq. 
The current implementation of the Demaq message store is built on the foundation of Natix
\cite{Fiebig:2003:ANX}, a native XML data store that is available as a C++ library.  While Natix allows for the
efficient, reliable and persistent storage of XML fragments, we had to extend 
 the existing collection-based storage and recovery subsystems with
recoverable queues for XML message storage
\footnote{The queue extensions (but not yet the Demaq Rule Engine) are included in Natix V2 available at http://db.informatik.uni-mannheim.de/natix.html.en}. Since the XML queues and the XML collections have similar storage formats,
we can leverage the Natix run-time system for rule execution. This includes the Natix virtual machine
for query evaluation, the recovery and locking subsystems as well as the schema management components.

Having a full-fledged XDS in source form, we can also apply existing database and transaction processing
research to message processing. For example, our append-only approach for message queues
simplifies logging and recovery because there are fewer in-place updates. 
Further, our declarative mechanism for specifying message retention (Sec.~\ref{sect:messageretention})
frees the system from the need to fully log message deletions -- after a crash, the
decision to delete certain messages can be reached without analyzing the log. 

However, Natix is not the only option as underlying message store. The Demaq design is
highly modular and could also be implemented on top of other existing,
queue-enabled database systems, such as the Microsoft SQL Server Service Broker
(MSSSB) \cite{Wolter:2005:MSS}. MSSSB includes some of the necessary features, such as
data retention and security, and even incorporates queue-based system facilities
closely related to our gateway and echo queues.  While MSSSB already allows to
create messaging-based applications, it uses the TSQL programming
language, which is imperative at the top level and requires much tedious glue code
to work with XML data. The "activation" service that can be used to
automatically run programs on message arrival does not include declarative conditions and hence
complicates automatic optimizations. 
However, it could be used to trigger Demaq's rule processing and scheduling. 

\subsection{Gateway Queues}

To interface with remote Web Services, Demaq includes a communication
subsystem, offering both synchronous and asynchronous communication channels to
application rules in form of gateway queues.  For synchronous calls, system properties are used to correlate
request and reply messages.
Demaq provides SOAP bindings to transport protocols such as
HTTP and SMTP and the underlying TCP/IP client-server functionality.

\subsection{Slicing}

While slicing can be implemented by merging the slice definition into the rules
(see below), this would require to evaluate a complex query for every incoming
message. Instead, similar to the materialized views concept in RDBMSs, it is
possible to maintain a physical representation of the slices, for example using
a B-Tree indexed by the slice key. 

Slices also present an opportunity for improving concurrency, as they
form a natural new granularity, coarser than messages, but orthogonal to
queues.  By locking just the affected  slices, full serializability of the
individual message-processing transactions can be guaranteed without locking
whole queues.

\subsection{Rule Processing}
\label{sec:ruleproc}
The rule-processing module of Demaq is responsible for executing the application logic defined by QML rules. Its main building blocks are a rule compiler and a scheduling
component enforcing our execution model.

\subsubsection{Rule Compiler}
On deployment of an application, the rule compiler is used to compile the application's
rule set into execution plans. 
Essentially, rule evaluation is a query against the state of the underlying storage engine
that results in an update list. Hence, we
can reuse execution plans and optimization techniques for XML query processing
to perform rule evaluation. 

For each queue, the compiler collects all rules that are associated with it.
The bodies of the rules represent XQuery Update expressions, which are
rewritten.  Rewriting includes supplying default parameters to functions which
depend on the current queue (such as \texttt{qs:queue()}). Similar to conventional view
merging, fixed properties (see Sec.~\ref{sec:props}) are inlined, as is slice
access that is not materialized (see above).  After rewriting, the rule bodies
are combined into a single query by  
concatenating all pending actions into a single sequence.  The query is then
compiled into an execution plan that is executed every time a message arrives
in that queue.  A variety of existing techniques can be leveraged to improve
processing performance, including XML filtering \cite{Diao:2003:QPX}, efficient
expression evaluation \cite{Yalamanchi:2003:MED}, and template folding
\cite{Kanne:2006:TFX}. 

The technique outlined above creates a "canonical" execution plan as a starting
point.  We intend to exploit the fully declarative nature of our language by
performing cost-based optimization by rewriting rule sets into equivalent, but
more efficient ones. 
 
\subsubsection{Scheduler}
Demaq's scheduling component implements the execution model introduced in Sec~\ref{sect:execution}.
The scheduler maintains a list of all unprocessed
messages and chooses the next message to be handled, considering both their
temporal ordering and the priority of the containing queues. Thus, a message in
a high priority queue may be processed before another one stored in a queue
with a lower priority, even if it has been created more recently.

The scheduler also orchestrates background tasks, such as gateway queue processing and
the message garbage collection for messages that do not have to be retained.

\section{Conclusion}
\label{sect:concl}
This paper discusses the Demaq approach for declarative XML message
processing.  Demaq may be used to specify the behavior of nodes in a variety
of distributed application scenarios, including, but not limited to Web Service implementation and
orchestration.  The major feature of Demaq is a fully declarative language that
is based on XQuery and has a simple semantics which facilitates optimization.
While implementing the Demaq execution model on top of
a queue-enabled DBMS such as our native XML data store Natix \cite{Fiebig:2003:ANX},
we found that we can benefit from the application of many existing techniques for
transaction processing and declarative query processing to message processing.

Much remains to be done. For example, time is an important aspect for Active
Web applications.  While our ordered queue model implies some relationship of
messages with respect to time, and we have mentioned time-based echo queues in
Sec.\ref{sect:timebased}, we have not explained in detail how time-based
conditions are incorporated into our language, and how they are evaluated.
Further, Demaq applications currently rely on a static set of queues, slicings,
and rule definitions that cannot be adapted during system runtime. As a result,
each time an application evolves, the processing system has to be shut down and
restarted. Clearly, this is unacceptable for zero-downtime environments,
bringing up the question how to allow for dynamic queue and rule evolution,
while still guaranteeing correct and reasonable system behavior.

\paragraph*{Acknowledgments}
We thank the anonymous referees for their helpful hints on improving this paper.
We also thank Simone Seeger and Matthias Brantner for their comments on the manuscript.

\bibliographystyle{plain}
\bibliography{../bib/custom}

\end{document}